\documentclass[acmsmall,screen,nonacm]{acmart}

\usepackage{tabularx}
\usepackage{multirow}
\usepackage{tabularray}
\UseTblrLibrary{booktabs}
\usepackage{listings}
\usepackage[normalem]{ulem}
\usepackage{xurl}
\usepackage[capitalise,nameinlink,noabbrev]{cleveref}

\setcopyright{none}
\acmDOI{}
\acmISBN{}
\makeatletter
\let\hyxmp@parse@acmart\relax
\makeatother

\newcolumntype{L}[1]{>{\raggedright\arraybackslash}p{#1}}
\definecolor{codebg}{RGB}{248,248,248}
\definecolor{codeframe}{RGB}{190,190,190}
\definecolor{codekw}{RGB}{0,82,155}
\definecolor{codeid}{RGB}{0,105,120}
\definecolor{codestring}{RGB}{148,82,0}
\definecolor{codecomment}{RGB}{85,110,85}
\definecolor{codekey}{RGB}{116,63,164}
\definecolor{codenum}{RGB}{100,100,100}

\lstdefinelanguage{Kotlin}{
  morekeywords={as,break,class,continue,do,else,false,for,fun,if,in,interface,
    is,null,object,package,return,super,this,throw,true,try,typealias,typeof,
    val,var,when,while,by,catch,companion,constructor,delegate,dynamic,
    file,finally,get,import,init,param,property,receiver,set,setparam,where},
  sensitive=true,
  morecomment=[l]{//},
  morecomment=[s]{/*}{*/},
  morestring=[b]"
}

\lstdefinelanguage{KnowledgeYAML}{
  sensitive=true,
  alsoletter={-},
  morecomment=[l]{\#},
  morestring=[b]",
  morekeywords={Evidence-backed,Bounded,Actionable,Knowledge,Summary,
    Applicable,Conditions,Fallback,Measures,Skippable,Content,Priority,
    Checks,Code,Evidence,Match,Matching,Rules,Downstream-code,
    Downstream-Code,Hints}
}

\lstdefinestyle{kotlinblock}{
  language=Kotlin,
  backgroundcolor=\color{codebg},
  rulecolor=\color{codeframe},
  keywordstyle=\color{codekw}\bfseries,
  emph={TextField,PasswordField,UiField,String,ConnectionConfig,
    settingResolver,protocolHandler,networkClient,resolve,connect,copy,
    authenticate},
  emphstyle=\color{codeid},
  commentstyle=\color{codecomment}\itshape,
  stringstyle=\color{codestring},
  numbers=left,
  numberstyle=\scriptsize\color{codenum},
  numbersep=5pt,
  stepnumber=1,
  xleftmargin=2.2em,
  xrightmargin=0.2em,
  framexleftmargin=1.1em,
  frame=single
}
\lstdefinestyle{knowledgeblock}{
  language=KnowledgeYAML,
  backgroundcolor=\color{codebg},
  rulecolor=\color{codeframe},
  keywordstyle=\color{codekey}\bfseries,
  numbers=none,
  breakautoindent=false,
  breakindent=0pt,
  frame=single
}

\crefname{listing}{Code Block}{Code Blocks}
\Crefname{listing}{Code Block}{Code Blocks}
\crefname{lstlisting}{Code Block}{Code Blocks}
\Crefname{lstlisting}{Code Block}{Code Blocks}

\providecommand{\rqanswer}[2]{%
\par
\noindent\colorbox{black!7}{%
\begin{minipage}{\dimexpr\columnwidth-2\fboxsep\relax}
\small\textbf{#1.} #2
\end{minipage}}
\par
}

\begin{document}

\title[FlowArk: Context-Aware Knowledge Reuse for Agentic Data-flow Analysis]{FlowArk: Boosting Agentic Data-flow Analysis for Android Apps via Context-Aware Knowledge Reuse}

\author{Yiming Zhang}
\email{zhangym253@mail2.sysu.edu.cn}
\affiliation{%
  \institution{Sun Yat-sen University}
  \country{China}
}

\author{Jiangrong Wu}
\email{jiangrongwu@cse.ust.hk}
\affiliation{%
  \institution{The Hong Kong University of Science and Technology}
  \country{China}
}

\author{Yuhong Nan}
\authornote{Corresponding author: Yuhong Nan (nanyh@mail.sysu.edu.cn).}
\email{nanyh@mail.sysu.edu.cn}
\affiliation{%
  \institution{Sun Yat-sen University}
  \country{China}
}

\renewcommand{\shortauthors}{Zhang, Wu, and Nan}

\begin{abstract}
Data-flow analysis is foundational to Android app privacy and security auditing. Recent coding agents can assist with non-trivial source-to-sink data-flow analysis tasks by searching, reading, and reasoning over repository code. However, when these tasks are executed as a batch workload, current agentic analysis setups incur substantial re-analysis cost. Agent instances assigned to different taint sources may inspect shared code fragments, because code reuse in the target app can cause different data-flow paths to converge on shared program logic. Since these agent instances are context-isolated, analysis of these shared code fragments can be repeated within a batch, unnecessarily consuming API budget and limiting scalability.

We propose FlowArk, a knowledge-reuse system that reduces re-analysis cost in batch agentic data-flow analysis by making knowledge from completed analyses available to later agent instances. Specifically, FlowArk distills completed analysis histories into reusable knowledge candidates, packages these candidates into matchable knowledge entries, and injects matched entries into a later agent instance's context. We implement FlowArk on OpenCode and evaluate it on 4,649 source-to-sink data-flow analysis tasks from 49 open-source Android apps. Compared with standard OpenCode, FlowArk-enabled OpenCode maintains comparable analysis quality while reducing end-to-end API cost by 26.91\%. This cost-reduction trend holds across different LLM backends and repeated runs.
\end{abstract}


\maketitle

\section{Introduction}

Large language models and coding agents are being used to support security analysis tasks such as data-flow analysis \cite{wang_llmdfa_2024,ghebremichael_multi-agent_2026}, repository-level security auditing \cite{guo_repoaudit_2025}, and vulnerability detection \cite{li_iris_2025,li_llm-based_2026,yildiz_benchmarking_2025}. A common characteristic of these systems is that coding agents can iteratively invoke search and code-reading tools within a repository, gradually collect evidence, reason about data-flows, and ultimately produce security analysis results. We refer to source-to-sink data-flow analysis performed in this agent-driven manner as \emph{agentic data-flow analysis}. This setting is attractive because coding agents can analyze complex cross-file data-flows and reduce the manual effort required to customize taint propagation rules.

In current Android program analysis tasks such as privacy compliance auditing and security vetting, agentic data-flow analysis is often organized as a batch workload. For example, an analyst may define hundreds of untrusted UI input sources and use agents to analyze whether tainted data from each source can reach security-sensitive sink APIs. These analysis tasks are often assigned to multiple agent instances, or independent sessions\footnote{For simplicity, we opt to use the term ``agent instance'' across the paper.}, to reduce task complexity and control the context size. Each agent instance is context-isolated and maintains its own reasoning trace.

\begin{figure}[t]
\centering
\includegraphics[width=0.65\linewidth]{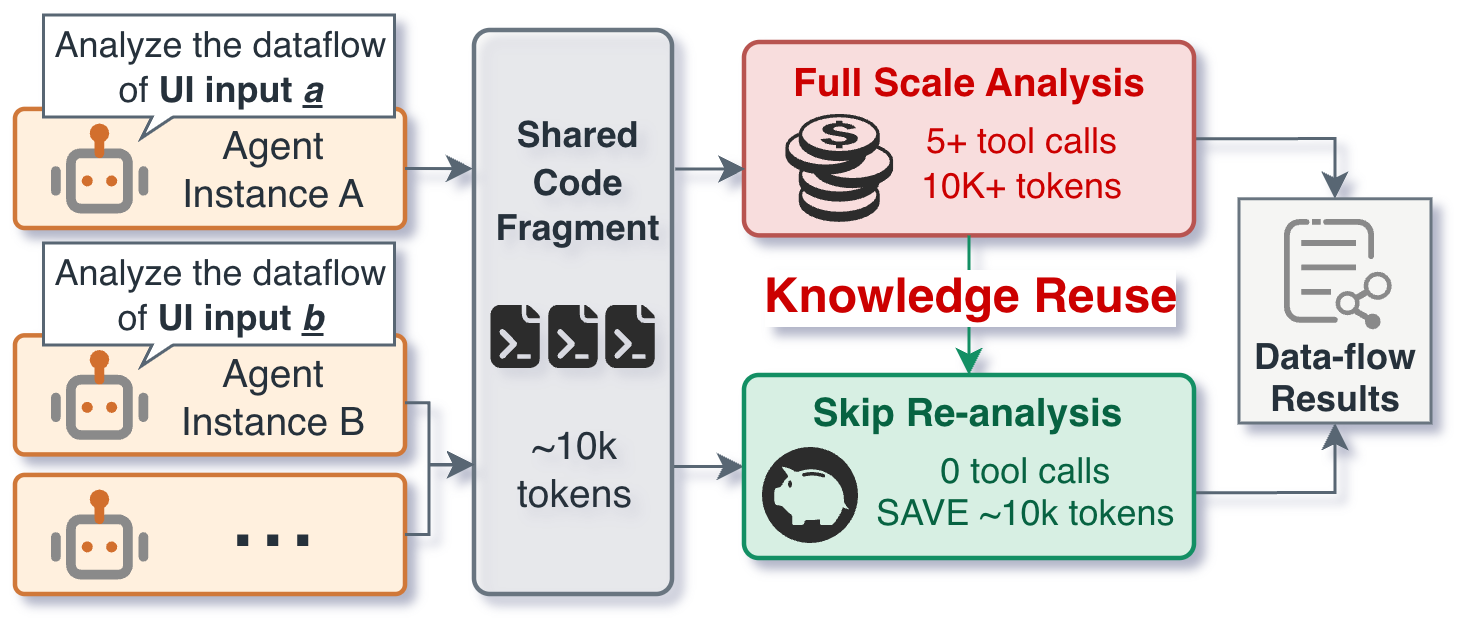}
\vspace{-0.5em}
\caption{The reuse opportunity in batch agentic data-flow analysis. Agent instances analyzing a shared code fragment can reuse prior analysis knowledge to avoid repeated reading and reasoning.}
\label{fig:reanalysis-cost}
\Description{Two agent instances analyze different UI input sources that converge on one shared code fragment. The diagram contrasts full reanalysis with skipping the shared fragment using knowledge from an earlier analysis.}
\end{figure}

This batch analysis setup involves substantial unnecessary re-analysis, which lowers efficiency and increases token consumption. Modern Android apps commonly encapsulate frequently used program logic, such as event dispatch, into shared components for code reuse. Consequently, data-flow analysis tasks that start from different sources can converge on the same shared code fragments (overlapping code regions relevant to multiple source-to-sink data-flows), causing later agent instances to re-analyze code that earlier instances have already inspected. Such shared code fragments often involve callbacks, event dispatch, or inter-component communication logic that requires cross-file, multi-step confirmation, making repeated analysis especially costly~\cite{cao_edgeminer_2015,li_iccta_2015,yan_comprehensive_2023}.

The root cause of this bottleneck is the lack of effective analysis knowledge reuse across context-isolated agent instances. Without access to an earlier agent instance's analysis knowledge, a later agent instance repeats similar tool calls and reasoning steps to re-analyze the same shared code fragment. Although traditional static program analysis has long reduced repeated analysis cost by reusing summaries, cached analysis results, or incremental updates~\cite{arzt_reviser_2014,cai_leveraging_2018,dusing_persisting_2023}, batch agentic data-flow analysis lacks an analogous cross-instance mechanism: prior code understanding remains inside the earlier instance's context and is unavailable when a later instance reaches the same shared code fragment. \Cref{fig:reanalysis-cost} illustrates this gap and reuse opportunity: an earlier instance may pay the full scale analysis cost for a shared code fragment, while a later instance could skip re-analysis if the prior analysis knowledge were available, reducing unnecessary token consumption.

Current general agent memory systems provide limited support for this data-flow-analysis-oriented knowledge reuse. General agent memory systems such as Mem0~\cite{mem0} are usually designed for broad coding assistance, such as summarizing previous interactions. In batch agentic data-flow analysis, they face two fundamental limitations: (1) \emph{Imprecise knowledge extraction.} General agent memory has difficulty precisely extracting the reusable knowledge needed for later data-flow analysis tasks. A completed analysis history contains verbose tool-call logs and task-specific data-flow results. Without a data-flow-analysis-oriented memory design, a general memory system may preserve imprecise or low-value parts of the analysis history. (2) \emph{Unreliable reuse timing.} General memory systems typically rely on the active agent instance to issue memory queries on its own, making reuse depend on whether the agent notices the opportunity in time. The reuse opportunity is usually a narrow window: it is not visible in the initial task description and emerges only after tool calls reveal files, method calls, or fields that overlap with previous analyses. As a result, agent-initiated retrieval may query memory only after the shared fragment has already been re-analyzed, or may miss the reuse opportunity entirely.

To this end, we propose FlowArk, a knowledge reuse system for batch agentic data-flow analysis. FlowArk helps context-isolated agent instances share reusable analysis knowledge so that later agent instances can reduce re-analysis cost. The core idea is to turn knowledge reuse from agent-initiated memory search into runtime-context-triggered, rule-based matching of reusable knowledge. FlowArk realizes this idea through a three-stage reuse loop. First, \emph{Selective Knowledge Distillation} uses overlaps among historical data-flow results to identify shared code fragments and distill reusable knowledge candidates. Second, \emph{Knowledge Packaging and Admission} packages these candidates with matching rules, allowing FlowArk to recognize when an active agent instance reaches a corresponding shared code fragment. Third, \emph{Context-Aware Runtime Knowledge Injection} monitors the active agent instance's runtime analysis context and injects compact matched knowledge entries to guide the agent to skip re-analysis.

We implemented FlowArk with OpenCode~\cite{opencode_docs} and evaluated it on 4,649 source-to-sink data-flow analysis tasks from 49 open-source Android apps. Compared with standard OpenCode, FlowArk reduces end-to-end LLM API cost by 26.91\%, corresponding to a throughput of 1,059 versus 774 tasks per USD 100. Further evaluations on a 1,284-task subset\footnote{We use a subset because these evaluations incur substantial API costs. \Cref{sec:eval-setup} details the selection procedure.} show that FlowArk achieves larger cost savings than two history-reuse baselines, \emph{Mem0-enabled OpenCode} and \emph{RAG-enabled OpenCode}, while maintaining comparable analysis quality. Moreover, this cost-reduction trend holds across different LLM backends and repeated runs.

This paper makes the following contributions:

\begin{itemize}
\item We characterize a re-analysis bottleneck in batch agentic data-flow analysis, where context-isolated agent instances repeatedly analyze shared code fragments and accumulate unnecessary token/API cost across a batch workload.
\item We propose FlowArk, a knowledge-reuse system for boosting batch agentic data-flow analysis. FlowArk distills reusable knowledge from completed analyses, packages it into matchable knowledge entries, and injects matched entries into the active agent context to reduce re-analysis cost.
\item We evaluate FlowArk on real Android app data-flow analysis workloads, showing consistent cost savings across LLMs and repeated runs while maintaining comparable analysis quality.
\end{itemize}

\section{Motivating Example and Design Challenges}
\label{sec:background-and-problem-statement}

\subsection{Motivating Example}
\label{sec:motivating-example}

\begin{figure}[!t]
\centering
\includegraphics[width=\linewidth]{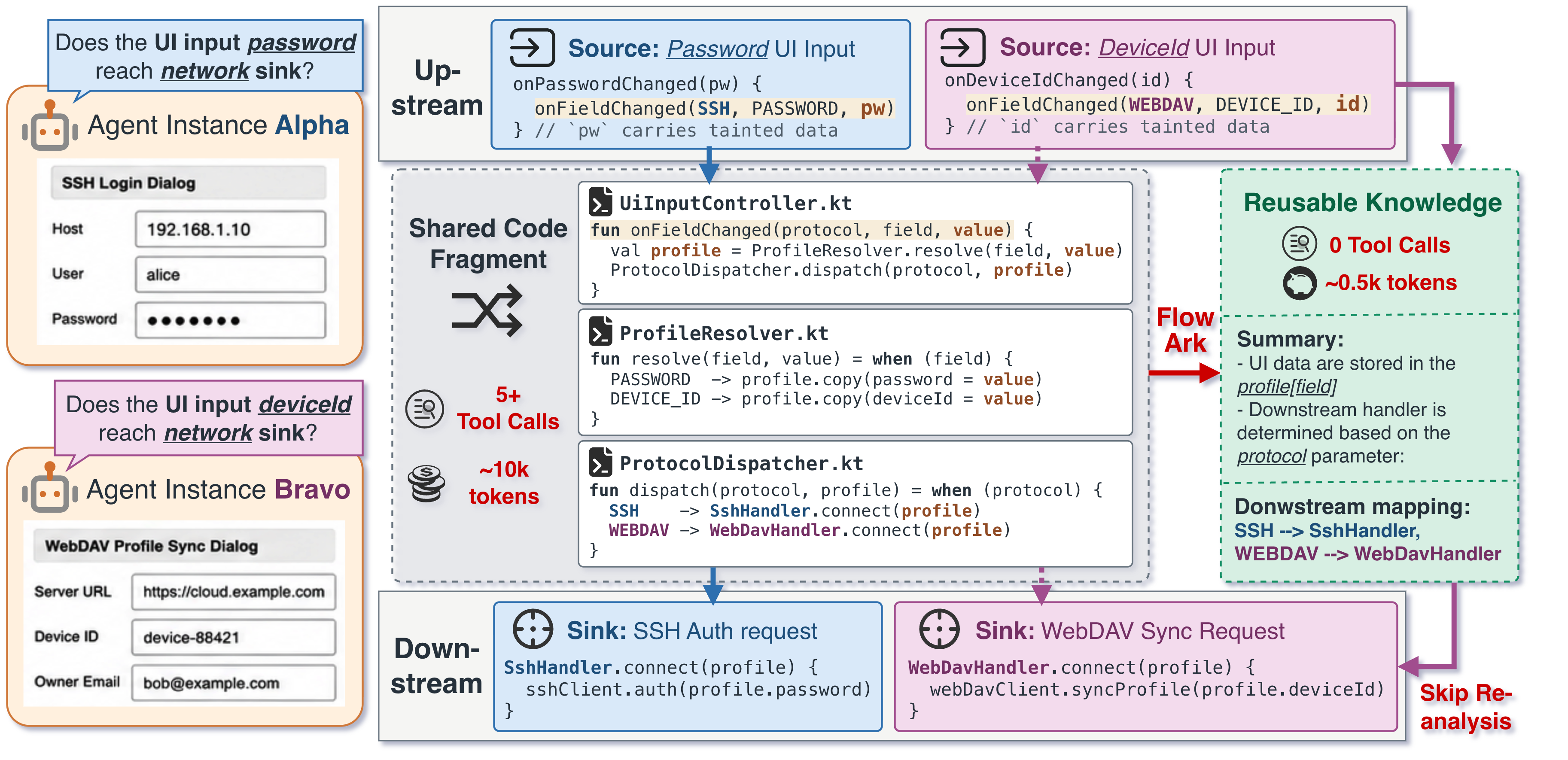}
\vspace{-1.7em}
\caption{Motivating example of knowledge reuse. Alpha and Bravo analyze different UI sources through shared middle-layer logic. Bravo reuses Alpha's knowledge to skip re-analysis of this shared code fragment.}
\label{fig:motivating-example}
\Description{Two source-specific analysis paths, Alpha and Bravo, converge on shared UI input handling and protocol dispatch, then continue to protocol-specific network handlers. A reusable knowledge block describes the common middle-layer mechanism and downstream branches.}
\end{figure}

\Cref{fig:motivating-example} demonstrates a concrete knowledge-reuse opportunity between two data-flow analysis tasks. Agent Instance Alpha analyzes whether the \texttt{password} UI input source can reach a network sink, and Agent Instance Bravo later analyzes the \texttt{deviceId} UI input source. The two UI callbacks pass different \texttt{protocol} and \texttt{field} values: Alpha uses \texttt{SSH}/\texttt{PASSWORD}, while Bravo uses \texttt{WEBDAV}/\texttt{DEVICE\_ID}. Their downstream sink checks also differ: Alpha eventually checks \texttt{SshHandler}, while Bravo needs to check \texttt{WebDavHandler}. Despite these upstream and downstream differences, both data-flows pass through the same shared code fragment: the middle-layer setting-resolution and protocol-handler-dispatch logic.

Alpha's analysis of this shared code fragment contains reusable knowledge that can reduce Bravo's re-analysis cost. Through cross-file analysis, Agent Instance Alpha can confirm two data-flow facts: the resolver copies the input \texttt{value} into the profile slot selected by the \texttt{field} parameter, and the dispatcher routes the resolved \texttt{profile} to the handler selected by the \texttt{protocol} parameter. This knowledge is not tied to Alpha's \texttt{SSH}/\texttt{PASSWORD} analysis task alone. When Agent Instance Bravo reaches the same shared code fragment while analyzing the \texttt{WEBDAV}/\texttt{DEVICE\_ID} task, this reusable knowledge remains applicable: \texttt{DEVICE\_ID} selects the \texttt{profile.deviceId} slot, and \texttt{WEBDAV} selects the WebDAV handler.

However, this reuse opportunity does not automatically produce a clean artifact, and its reuse window is narrow. Alpha's completed history contains SSH-password-specific reasoning, while Bravo needs bounded middle-layer knowledge and timing guidance that lets it skip re-analysis before the opportunity is lost. This example gives a concrete target for FlowArk: extract the reusable middle-layer analysis knowledge from Alpha, make it recognizable when Bravo reaches the same middle-layer, and inject it early enough for Bravo to avoid re-analysis.

\subsection{Design Challenges and Solutions}
\label{subsec:design-challenges-and-solutions}

The motivating example highlights two requirements for knowledge reuse. First, reusable analysis knowledge needs to be precisely extracted from verbose, task-specific analysis histories. Second, matched knowledge must reach later agent instances when their runtime analysis context reveals a reuse opportunity. These requirements lead to the following three design challenges.

\noindent \textbf{\textit{C1: How to distill reusable knowledge from completed analysis histories?}} The first challenge is to extract reusable analysis knowledge from a completed data-flow analysis history while filtering out irrelevant task-specific details. In \Cref{fig:motivating-example}, Alpha's analysis history mixes the reusable resolver-dispatcher mechanism with task-specific details such as the \texttt{password} source, the \texttt{SSH} protocol value, and the SSH authentication sink API. Bravo needs bounded knowledge of the shared mechanism that explains which middle-layer analysis can be skipped and which downstream checks remain necessary.

FlowArk addresses this challenge through \emph{Selective Knowledge Distillation}. After a data-flow analysis task completes, FlowArk asks the same agent instance to generate knowledge candidates under three selection-rule constraints: evidence-backed, bounded, and actionable. These constraints make each candidate traceable to code evidence, limited to a shared code fragment, and useful for deciding what later agent instances can skip and where they should continue. FlowArk also uses overlaps among historical data-flow analysis results to identify shared code fragments. In the motivating example, the shared resolver-dispatcher segment in Alpha's and Bravo's paths illustrates the shared code fragment that knowledge extraction should cover.

% \subsubsection{\textbf{How to Package Knowledge for Runtime Matching}}
\noindent \textbf{\textit{C2: How to package knowledge for runtime matching?}}
The second challenge is to make each knowledge candidate matchable by associating it with stable code anchors of the corresponding shared code fragment. In \Cref{fig:motivating-example}, Bravo's initial data-flow analysis task only states that \texttt{deviceId} should be checked against a network sink. This initial task description does not reveal that Bravo will enter the same resolver-dispatcher mechanism analyzed by Alpha. The reuse opportunity becomes visible only when Bravo's code-reading tool outputs expose code cues such as \texttt{UiInputController.onFieldChanged}. Consistent with prior work on structured memory~\cite{wu_repoformer_2024,wang_memgovern_2026,wang_repomem_2026}, a robust reuse mechanism should store each candidate with explicit matching rules over such anchors, so later runtime contexts can trigger the corresponding reusable knowledge.

FlowArk addresses this challenge through \emph{Knowledge Packaging and Admission}. It packages each distilled candidate into a matchable knowledge entry with matching rules over stable code anchors, including symbol paths, method calls, field names, and package scopes. These rules describe the runtime analysis context in which the knowledge becomes relevant. In \Cref{fig:motivating-example}, a knowledge entry can use \texttt{UiInputController.onFieldChanged(protocol, field, value)} as a matching anchor and record protocol-to-handler mappings such as \texttt{SSH} to \texttt{SshHandler} and \texttt{WEBDAV} to \texttt{WebDavHandler}. Before storage, FlowArk admits only knowledge entries with specific matching rules, sufficient evidence, clear boundaries, and actionable guidance. Admitted entries are then ready for later runtime matching.

% \subsubsection{\textbf{How to Inject Matched Knowledge for Cost Reduction}}
\noindent \textbf{\textit{C3: How to inject matched knowledge for cost reduction?}}
The third challenge is to inject matched knowledge at runtime early enough to prevent re-analysis of the shared code fragment. Once Bravo's tool outputs expose matching anchors such as \texttt{onFieldChanged(WEBDAV, DEVICE\_ID, id)}, the system can recognize that Alpha's resolver-dispatcher knowledge applies. However, the match is useful only if the knowledge enters Bravo's context before Bravo spends more tool calls re-analyzing the shared code fragment.

FlowArk addresses this challenge through \emph{Context-Aware Runtime Knowledge Injection}. During an active agent instance, FlowArk checks admitted knowledge entries after new tool outputs arrive. It applies each entry's matching rules to the newly exposed runtime analysis context. When Bravo's runtime analysis context satisfies the matching rules for the resolver-dispatcher knowledge, FlowArk injects a compact block that summarizes the reusable mechanism, indicates the skippable analysis steps, and selects only the downstream hints relevant to the current analysis context. This design turns the matching rules prepared by \emph{Knowledge Packaging and Admission} into timely runtime reuse: knowledge is delivered after enough evidence supports applicability and before re-analysis begins.

\section{FlowArk Overview}

\begin{figure}[thbp]
\centering
\includegraphics[width=\linewidth]{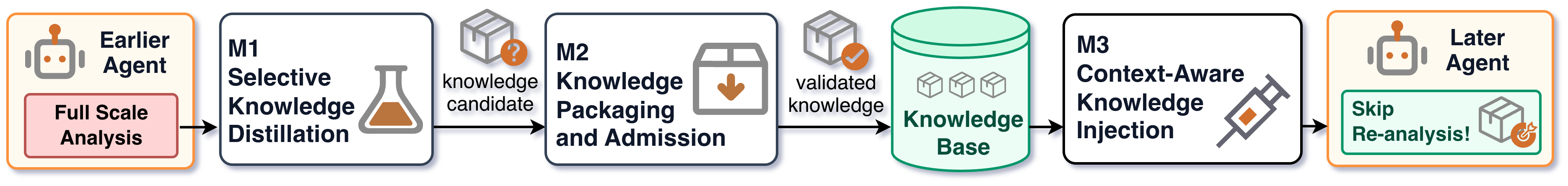}
\vspace{-1.5em}
\caption{Workflow of FlowArk.}
\label{fig:flowark-overview}
\Description{A completed analysis feeds selective knowledge distillation and knowledge packaging and admission. Admitted entries enter the knowledge base, which supplies context-aware runtime knowledge injection to later analyses.}
\end{figure}

\Cref{fig:flowark-overview} shows how FlowArk reuses analysis knowledge across context-isolated agent instances. FlowArk runs alongside standard agent instances without changing their implementation. Once an agent instance completes a data-flow analysis task, FlowArk takes its analysis history as input and starts the knowledge reuse pipeline. First, \emph{Selective Knowledge Distillation} extracts knowledge candidates about shared code fragments from the analysis history. Then, \emph{Knowledge Packaging and Admission} checks these candidates, attaches matching rules, and stores admitted candidates as knowledge entries in the knowledge base.

While an agent instance is active, FlowArk uses admitted knowledge entries to help it avoid re-analysis of matched shared code fragments. The \emph{Context-Aware Runtime Knowledge Injection} module observes the runtime analysis context exposed by the agent's tool outputs. When this context matches an admitted knowledge entry, FlowArk assembles a compact injection block for the matched knowledge. The block enters the agent context before the agent proceeds into re-analysis, allowing the agent instance to skip the matched shared code fragment and continue the downstream checks for the current data-flow analysis task.

\section{Design Details}

This section details how FlowArk implements the three modules in \Cref{fig:flowark-overview}.

\subsection{Selective Knowledge Distillation}

\label{sec:distillation}

This module converts a completed data-flow analysis task history into several reusable knowledge candidates. To perform this conversion, FlowArk sends a follow-up distillation request to the same agent instance that completed the analysis; the module first constrains this request with selection rules for evidence-backed, bounded, and actionable candidates, then mines recurring method-call-chain fragments across historical analyses to guide candidate boundaries.

\subsubsection{\textbf{Post-Analysis Knowledge Distillation}}

\label{sec:post-analysis-distillation}

After a data-flow analysis task completes, FlowArk appends a structured distillation request that asks the same agent instance to extract reusable knowledge candidates under the three selection rules in \Cref{tab:selection-rules}. The request restricts the agent to code evidence already read during the completed analysis and forbids new code search, keeping distillation grounded while limiting additional exploration cost.

\begin{table}[htbp]
\centering
\caption{Selection rules for reusable knowledge candidates.}
\vspace{-0.5em}
\label{tab:selection-rules}
\fontsize{8.5pt}{10pt}\selectfont
\setlength{\tabcolsep}{3pt}
\renewcommand{\arraystretch}{1.25}

\begin{tabular}{>{\centering\arraybackslash}m{0.20\linewidth}m{0.75\linewidth}}
\toprule
\textbf{Selection Rule} & \textbf{Description} \\
\midrule
Evidence-backed &
Grounded in code evidence already read by the completed agent instance. \\
Bounded &
Shared fragment with a clear entry, boundary, applicable conditions, and fallback. \\
Actionable &
Specifies skippable code and downstream locations to inspect first.\\
\bottomrule
\end{tabular}
\end{table}

To apply these constraints to distillation, FlowArk encodes the selection rules as a six-field schema for every knowledge candidate. \emph{Knowledge Summary}, \emph{Applicable Conditions}, and \emph{Fallback Measures} define the candidate boundary; \emph{Skippable Content} and \emph{Priority Checks} make the candidate actionable for later analyses; and \emph{Code Evidence} grounds the candidate in files or code snippets.

\Cref{fig:knowledge-candidate-example} instantiates this six-field schema for the motivating example. The candidate records the field-to-profile and protocol-to-handler mechanism, states when the knowledge applies, and tells a later analysis where to continue after skipping the shared code fragment.

\begin{figure}[tbp]
\centering
\begin{minipage}[t]{0.52\linewidth}
\begin{lstlisting}[style=knowledgeblock,basicstyle=\footnotesize\ttfamily]
Knowledge Summary: `field` selects the profile slot to update; `protocol` selects the downstream network handler to inspect.
Applicable Conditions: Use when the active path reaches `UiInputController.onFieldChanged(protocol, field, value)`.
Fallback Measures: If this callback is bypassed, read the alternate path.
Skippable Content: Skip field-to-profile assignment and protocol dispatch.
Priority Checks: Continue at the handler selected by the `protocol` parameter.
Code Evidence: UiInputController.kt@52: `onFieldChanged(...)`
\end{lstlisting}
\vspace{-0.7em}
\caption{Simplified knowledge candidate distilled from the motivating example.}
\label{fig:knowledge-candidate-example}
\Description{A YAML-style knowledge candidate lists its summary, applicable conditions, fallback measures, skippable content, priority checks, and supporting code evidence for UI input handling and protocol dispatch.}
\end{minipage}\hfill
\begin{minipage}[t]{0.45\linewidth}
\begin{lstlisting}[style=knowledgeblock,basicstyle=\footnotesize\ttfamily,
breakatwhitespace=true,breakindent=3em,
alsoletter={_,-},
emph={require_any,exclude},emphstyle=\color{codekw}\bfseries,
emph={[2]kind,receiver,method,value},emphstyle={[2]\color{codekw}\bfseries}]
Match Rules:
require_any:
  - kind: exact_symbol
    value: UiInputController.onFieldChanged
  ...
exclude:
  - kind: package_prefix
    value: com.example.test
Downstream-Code Hints:
  SSH --> Handlers.kt@100: SshHandler.connect
  WEBDAV --> Handlers.kt@273: WebDavHandler.connect
\end{lstlisting}
\vspace{-0.7em}
\caption{Matching rules and downstream-code hints attached to the candidate body in \Cref{fig:knowledge-candidate-example}.}
\label{fig:packaged-entry-example}
\Description{The packaged entry adds a positive symbol rule, a test-package exclusion rule, and downstream code hints that map SSH and WebDAV to their corresponding handlers.}
\end{minipage}
\end{figure}

\subsubsection{\textbf{Historical Overlap-Guided Boundary Selection}}
\label{sec:historical-overlap-boundary-selection}

FlowArk derives boundary guidance for knowledge distillation by comparing the current data-flow analysis result with historical results from the batch workload. A completed data-flow analysis result can contain a long source-to-sink data-flow path, so the selection rules alone may leave several plausible candidate spans. Some spans may be too broad to reuse efficiently, while others may be too narrow. Subpaths that recur across analysis results provide an additional signal for selecting among these candidate spans.

\begin{figure}[t]
\centering
\includegraphics[width=0.85\linewidth]{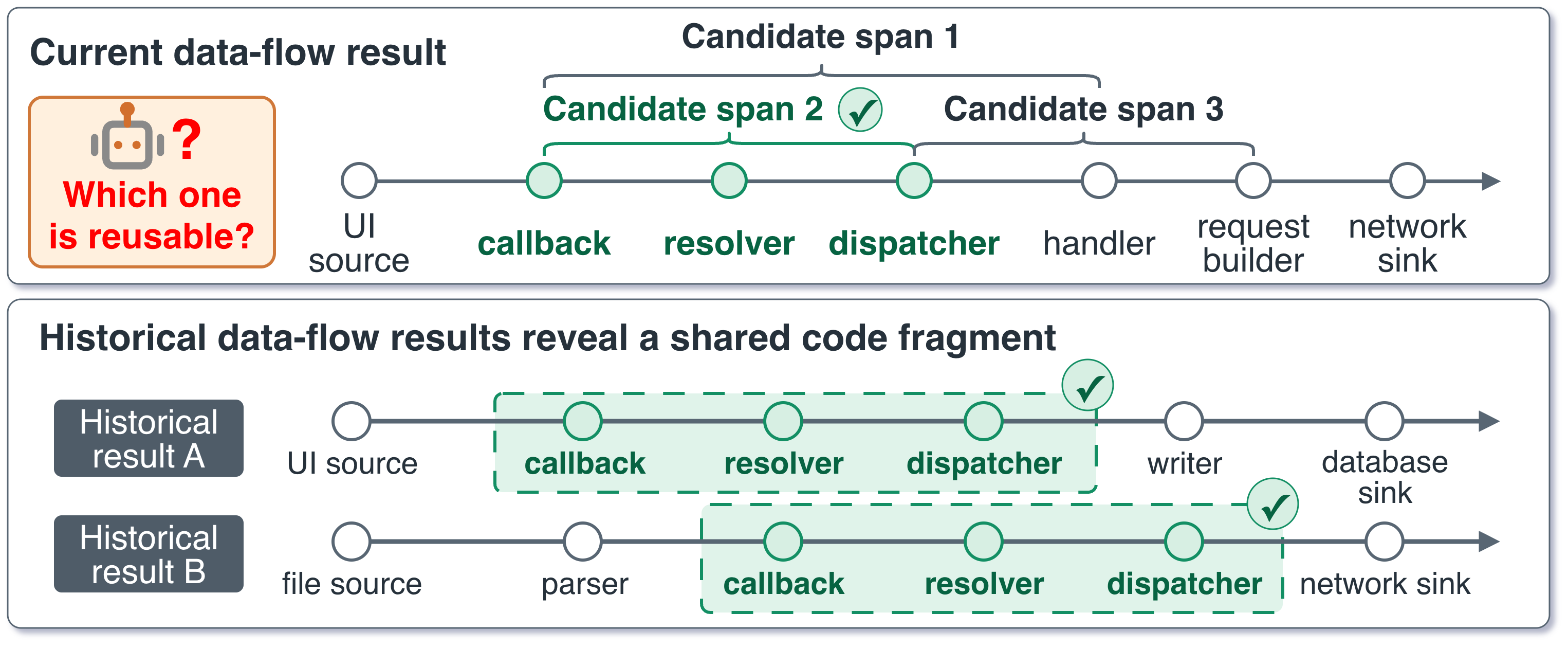}
\vspace{-0.7em}
\caption{Historical Overlap-Guided Boundary Selection. FlowArk compares the current data-flow result with historical results, then uses the recurring subpath as a boundary hint for knowledge candidate distillation.}
\label{fig:historical-overlap-boundary-selection}
\Description{Several candidate spans are shown within the current source-to-sink path. Two historical paths share the callback, resolver, and dispatcher segment, identifying a reusable boundary between source-specific and sink-specific code.}
\end{figure}

To identify these recurring subpaths, FlowArk extracts method call chains from each structured data-flow result. When a fragment of the current call chain also appears in historical data-flow results, FlowArk records it as a \emph{recurring subpath}, i.e., a method-call-chain fragment observed across multiple completed analyses.

Recurring subpaths turn historical overlap into a boundary signal for distillation. As shown in \Cref{fig:historical-overlap-boundary-selection}, the current and historical results share a \texttt{callback--resolver--dispatcher} subpath despite different sources and sinks. FlowArk uses this signal to bound the candidate around the reusable middle-layer mechanism.

\subsection{Knowledge Packaging and Admission}

\label{sec:packaging}

This module converts knowledge candidates produced by \emph{Selective Knowledge Distillation} into matchable knowledge entries and filters low-quality entries before storing them in the knowledge base. For this purpose, the agent instance that produces each knowledge candidate also proposes matching rules and downstream-code hints. FlowArk then attaches these proposed fields to the candidate and routes the resulting entry through an admission-and-repair step for final checks and targeted fixes before storage.

\subsubsection{\textbf{Knowledge Packaging}}

This stage packages each knowledge candidate into a knowledge entry that can be matched against runtime analysis context. Specifically, FlowArk attaches two proposed components to the candidate body: \emph{matching rules}, which describe when the entry becomes relevant, and \emph{downstream-code hints}, which describe where the active analysis should continue after the shared code fragment has been skipped.

\begin{table}[t]
\centering
\caption{Runtime matching rule types for knowledge entries.}
\vspace{-0.5em}
\label{tab:matching-rules}
\footnotesize
\setlength{\tabcolsep}{2pt}

\begin{tabularx}{\linewidth}{@{}>{\raggedright\arraybackslash}p{0.23\linewidth}>{\raggedright\arraybackslash}X@{}}
\toprule
\textbf{Rule Type} & \textbf{Example} \\
\midrule
\texttt{exact\_symbol} &
\texttt{com.example.ui.UiInputController.}\allowbreak\texttt{onFieldChanged} \\
\texttt{call} &
\texttt{receiver: UiInputController, method: onFieldChanged} \\
\texttt{symbol\_tail} &
\texttt{UiInputController} \\
\texttt{package\_\allowbreak{}prefix} &
\texttt{com.example.ui} \\
\bottomrule
\end{tabularx}
\end{table}

\Cref{tab:matching-rules} shows four matching rules with different levels of specificity. FlowArk matches each rule's code anchor directly against text from the runtime analysis context, keeping the checks lightweight. For example, a \texttt{call} rule checks for a method call with the specified receiver and method name.

Individual rule types describe how one code anchor is matched; FlowArk then organizes these anchors into three buckets to express how the anchors should be combined. \texttt{require\_any} contains individually sufficient positive anchors; \texttt{require\_all} combines anchors that must be observed together, such as a generic API anchor plus a concrete class or key. \texttt{exclude} suppresses entries in known inapplicable contexts. These buckets express both permissive and restrictive matching conditions without requiring a complex program-analysis query.

Matching rules select knowledge entries for injection; downstream-code hints guide subsequent analysis after reuse. A downstream-code hint is an optional branch-level annotation. It maps a key in the current runtime analysis context, such as a protocol value or receiver object, to a downstream code location. This supports parameter-sensitive or base-object-sensitive branch choices. In the motivating example, the \texttt{protocol} argument determines which network handler to check after skipping re-analysis of the shared code fragment: \texttt{SSH} maps to \texttt{SshHandler} and \texttt{WEBDAV} to \texttt{WebDavHandler}. FlowArk injects only the hints matching the current runtime analysis context.

\Cref{fig:packaged-entry-example} shows only the fields attached during packaging. A stored entry combines the candidate body represented by \Cref{fig:knowledge-candidate-example} with these matching rules and downstream-code hints.

\subsubsection{\textbf{Knowledge Admission and Repair}}

This stage prevents low-quality knowledge entries from entering the knowledge base. A poorly grounded entry may inject unsupported guidance, and an overly broad matching rule may trigger the entry in unrelated runtime analysis contexts. To reduce these risks, FlowArk applies admission checks to the candidate structure and matching rules before storage. Knowledge entries that pass admission are stored in the knowledge base. Knowledge entries with repairable rule-level issues enter a restricted repair step and then re-enter the same admission checks.

The admission check applies deterministic structural and matching-rule checks guided by \Cref{tab:selection-rules}. First, FlowArk checks for the required guidance fields describing skippable content, priority checks, and fallback conditions. Next, FlowArk checks whether the matching rules refer to code anchors that can be matched in the codebase and are specific enough for runtime matching. If a knowledge entry only lacks sufficiently strong rule anchors, FlowArk asks an agent instance to add or refine matching rules under a restricted budget. This repair step is limited to rule-level fixes; it does not ask the agent instance to add new code evidence or rewrite the candidate body. After repair, FlowArk reruns the same admission checks. Knowledge entries that still fail these checks are discarded.

Only admitted entries are available for runtime matching and injection.

\subsection{Context-Aware Runtime Knowledge Injection}

\label{sec:runtime-injection}

\begin{figure}[t]
    \centering
    \includegraphics[width=0.65\linewidth]{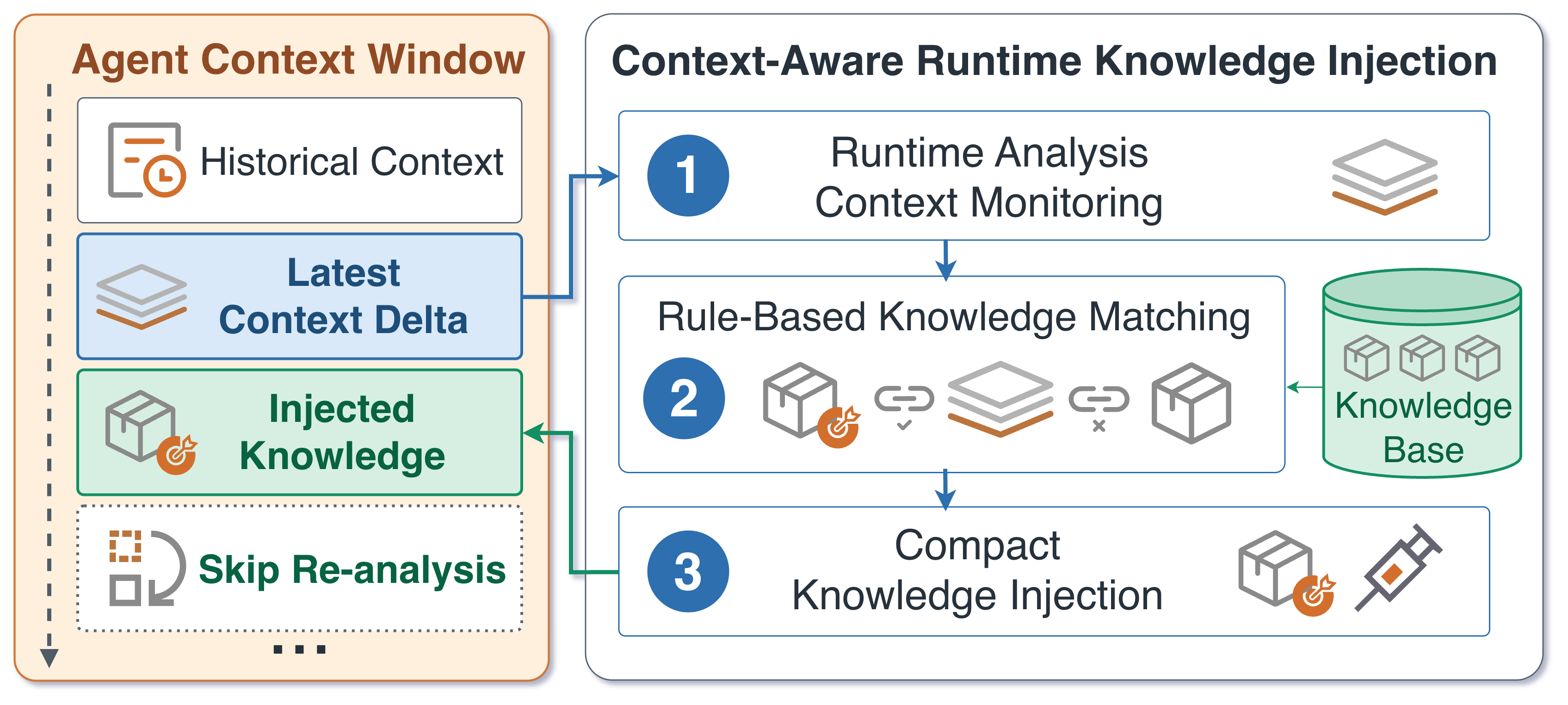}
    \vspace{-0.7em}
    \caption{Runtime knowledge injection process. FlowArk matches the latest context delta against knowledge, and injects relevant knowledge into the active agent instance's context to help the agent skip re-analysis.}
    \label{fig:runtime-injection}
\Description{The latest context delta from the agent is monitored, matched against knowledge-base rules, and used to select a compact knowledge injection before analysis continues.}
\end{figure}

This module injects relevant knowledge at runtime while limiting added context overhead. FlowArk matches admitted entries against the \emph{latest context delta}, the newly exposed text at a runtime boundary, and injects matched knowledge as a compact block before the agent continues analysis. \Cref{fig:runtime-injection} summarizes the process.

\subsubsection{\textbf{Runtime Analysis Context Monitoring}}

This stage determines when FlowArk should attempt knowledge matching. A useful matching point should provide enough code-level evidence to identify a shared code fragment and occur early enough for matched knowledge to guide the next analysis step. FlowArk therefore observes two stable runtime boundaries: first, at the analysis request submission, it uses the task description, source, and target sink category as matching text; second, after tool calls complete, it uses newly returned tool outputs, which can expose concrete files, method calls, or fields. The matching text collected at the current boundary forms the \emph{latest context delta}, which serves as the input to rule-based knowledge matching.

\subsubsection{\textbf{Rule-Based Knowledge Matching}}

This stage selects admitted knowledge entries whose matching rules are satisfied by the latest context delta. For each admitted knowledge entry, FlowArk evaluates the matching rules attached during \emph{Knowledge Packaging and Admission} in two steps: it begins by applying \texttt{exclude} rules, because a matched negative anchor indicates that the entry should not be used in the current context; for entries that remain, it evaluates \texttt{require\_all} and \texttt{require\_any} as positive matching evidence.

An entry is passed to the injection stage only when it has positive matching evidence and no satisfied exclusion rule in the latest context delta. The agent is instructed to skip re-analysis only when the entry's applicable conditions hold.

\subsubsection{\textbf{Compact Knowledge Injection}}

This stage selects how much matched knowledge to inject, assembles it into a compact block, and injects the block into the active agent's context. To guide selection, FlowArk ranks entries by scores derived from their rule matches. It also consults the agent instance's injection history to avoid repeated full-text injection. An entry matched for the first time may be injected with its main content within the character budget, whereas a previously injected entry may be represented by a short reminder or omitted. To further limit distractions and context overhead, FlowArk includes only downstream-code hints matching the latest context delta.

The assembled block is then inserted into the agent context through the available agent interface. In our OpenCode-based implementation, FlowArk inserts the block through the user-input channel; agent frameworks with hook support can also insert the same block after tool-call completion. In both cases, the injected knowledge enters the context before the agent continues its next analysis step.

\noindent\textbf{Method summary.}
Together, these modules form a reuse pipeline from completed analysis histories to runtime assistance for later agent instances. We next evaluate whether this design reduces re-analysis cost while preserving data-flow analysis quality.

\section{Evaluation}
\label{sec:evaluation}

We evaluate FlowArk through four research questions.

\begingroup
\begin{itemize}
    \item \textbf{RQ1: Overall efficiency.} To what extent does FlowArk reduce end-to-end token/API cost?
    \item \textbf{RQ2: Stability.} How consistently does FlowArk reduce cost across repeated runs, LLM backends, and different apps?
    \item \textbf{RQ3: Comparison with alternative mechanisms.} How does FlowArk compare with standard OpenCode and generic history-reuse mechanisms in cost and analysis quality?
    \item \textbf{RQ4: Sources of cost savings.} What contributes to FlowArk's observed cost savings?
\end{itemize}
\endgroup

We first describe the experimental setup and then answer the four research questions in \Crefrange{sec:eval-efficiency}{sec:eval-ablation}.

\subsection{Experimental Setup}
\label{sec:eval-setup}

\textbf{Dataset.}
Established Android benchmarks provide valuable controlled cases and manually validated real-app data-flows~\cite{arzt_flowdroid_2014,luo_taintbench_2022,mordahl_impact_2021}, but our evaluation of knowledge reuse also calls for workloads representative of data-flow analysis on contemporary apps. For example, TaintBench's explicitly declared target API levels range from 7 to 22, reflecting its curated apps' earlier platform settings.\footnote{The range is computed from explicit \texttt{targetSdkVersion} declarations in the released APKs: \url{https://github.com/TaintBench/TaintBench/releases/tag/TaintBenchSuite}.}

We therefore collect apps from F-Droid, a repository of thousands of open-source Android apps~\cite{fdroid_categories}. To cover diverse application types, we collect the first three apps displayed in each of the 72 official F-Droid categories. After deduplication by URL, the collection contains 171 apps. We download the latest available source release for each app before running any evaluated system.

We construct one data-flow analysis task for each source occurrence retained by predefined discovery and deduplication rules. The discovery rules locate API calls that read data and callback parameters that receive input. The retained sources cover five families: persistent storage reads, UI inputs, ICC payloads, Android platform APIs, and network payloads. Matches are then filtered using variable names and Android manifest declarations, with test and generated code excluded. After deduplication, each retained source occurrence is grounded in a concrete file, code line, statement, and source family. For each task, we pair the retained source with six shared target sink categories: \texttt{log}, \texttt{network}, \texttt{icc}, \texttt{file}, \texttt{database}, and \texttt{storage}.

\emph{From these tasks, we construct two workloads to balance experimental cost with realistic application sizes and task volumes.}

We use \textsc{Main49} to evaluate FlowArk's overall effectiveness on large batches of data-flow analysis tasks. We select the 50 apps with the largest source inventories and remove one duplicate app, yielding 49 distinct apps. The workload contains 4,649 data-flow analysis tasks.

\textsc{Sub15} supports the more costly experiments used to assess FlowArk's design and explain its cost savings. We randomly select 15 apps from \textsc{Main49} for repeated runs, baseline comparisons, comparisons across LLM backends, and module ablations. \textsc{Sub15} contains 1,284 tasks and retains all five source families.

\Cref{tab:eval-dataset} summarizes workload sizes and source-family distributions.

\begin{table}[!tbp]
\centering
\caption{Workload statistics and source distributions.}
\vspace{-0.5em}
\label{tab:eval-dataset}
\footnotesize
\setlength{\tabcolsep}{3pt}
\begin{tabular*}{\linewidth}{@{\extracolsep{\fill}}lrr*{5}{r}@{}}
\toprule
\multirow{2}{*}{\textbf{Workload}} & \multirow{2}{*}{\textbf{Apps}} & \multirow{2}{*}{\textbf{Tasks}} & \multicolumn{5}{c}{\textbf{Source family}} \\
\cmidrule(lr){4-8}
 & & & \textbf{Storage} & \textbf{UI} & \textbf{ICC} & \textbf{Platform} & \textbf{Network} \\
\midrule
\textsc{Main49} & 49 & 4,649 & 2,284 (49.1\%) & 1,074 (23.1\%) & 484 (10.4\%) & 342 (7.4\%) & 465 (10.0\%) \\
\addlinespace[2pt]
\textsc{Sub15} & 15 & 1,284 & 592 (46.1\%) & 308 (24.0\%) & 119 (9.3\%) & 149 (11.6\%) & 116 (9.0\%) \\
\bottomrule
\end{tabular*}
\end{table}

\textbf{Experimental configuration.}
We evaluate FlowArk using the workloads, systems, and LLM backends summarized in~\Cref{tab:eval-configuration}. The configurations of Standard OpenCode and FlowArk-enabled OpenCode are introduced in RQ1 (\Cref{sec:eval-efficiency}). RQ3 (\Cref{sec:eval-baselines}) describes the Mem0 and RAG history-reuse baselines, and RQ4 (\Cref{sec:eval-ablation}) describes the w/o M1--M3 module ablations.

\begin{table}[!htbp]
\centering
\caption{Experimental settings and objectives.}
\vspace{-0.5em}
\label{tab:eval-configuration}
\fontsize{8pt}{9.5pt}\selectfont
\begin{tblr}{
  width=\linewidth,
  colspec={Q[l,m] X[0.9,l,m] X[1.2,l,m] X[1.1,l,m]},
  colsep=3pt,
  column{1}={leftsep=0pt},
  column{4}={rightsep=0pt},
  rowsep=1pt,
  cell{3}{1}={r=2}{},
}
\toprule
\textbf{Workload} & \textbf{Systems} & \textbf{Backend(s)} & \textbf{Purpose} \\
\midrule
\textsc{Main49} & Standard, FlowArk & GLM-4.7 & Overall \& app-level cost reduction; knowledge reuse statistics \\
\midrule
\textsc{Sub15} & Standard, FlowArk & \mbox{GLM-4.7, GLM-5.3}, \mbox{DeepSeek-V4-Flash, MiniMax-M3} & {Cross-LLM-backend comparison;\\run-to-run stability$^{*}$;\\\mbox{contribution of knowledge reuse$^{*,\dagger}$}} \\
\cmidrule{2-4}
 & \mbox{Standard, FlowArk}, \mbox{Mem0, RAG}, w/o~M1--M3 & GLM-4.7 & {Baseline and ablation comparisons;\\\mbox{manual analysis quality audit$^{\ddagger}$}} \\
\bottomrule
\end{tblr}
\par\smallskip
\raggedright\footnotesize
$^{*}$These experiments were conducted later with GLM-5.3 because GLM-4.7 was no longer available through our LLM service.\\
$^{\dagger}$314 tasks sampled at fixed intervals span early, middle, and late knowledge accumulation stages.\\
$^{\ddagger}$257 tasks sampled within each app using deterministic SHA-256 hash ordering.
\end{table}

Across these configurations, all systems receive identical inputs for each task. Standard OpenCode uses the official tools~\cite{opencode_docs} to search and read code without write access, then reasons about and reports data-flows. Each task is assigned to a separate agent instance with its own isolated context. For each app, tasks follow the same sequence across the evaluated systems, and history reuse is restricted to earlier completed tasks in that app. FlowArk-enabled OpenCode adds only the prompts and mechanisms required for knowledge reuse; all other analysis instructions, tool permissions, and execution settings remain identical to those of standard OpenCode. \Cref{fig:analysis-prompt} shows the common analysis prompt excerpts.

% Sources: flowark/prompts/start_analysis-v3.md and opencode-adapter/flowark_system-v1.md.
\begin{figure}[tbp]
\centering
\begin{minipage}{0.98\linewidth}
\begin{lstlisting}[style=knowledgeblock,language={},breakatwhitespace=true]
Source: {source}; target sink types: {sink_types}
Trace the tainted data from the source until a sink, flow break, or irrelevant branch is confirmed. After finding a source-sink path, inspect its unchecked branches. At framework boundaries, search for downstream uses or callbacks. Prioritize paths strongly related to the source and sinks. Preserve key evidence and reasons for branch selection. Inspect code without modifying it and collect relevant data-flow evidence.
\end{lstlisting}
\vspace{-0.7em}
\caption{Excerpts of analysis input used in the experiment.}
\label{fig:analysis-prompt}
\Description{The prompt excerpt specifies the source and target sinks, flow tracing and branch handling, and evidence collection.}
\end{minipage}
\end{figure}

\textbf{Metrics.}
We use input tokens, ReAct turns, and API cost as complementary indicators of analysis effort, reflecting cumulative context volume, iterative execution, and monetary expenditure, respectively. Input-token and ReAct-turn counts provide a pricing-independent view of analysis effort. Unless otherwise stated, both metrics cover the full workflow, including knowledge management. We omit wall-clock time because the experiments span a long period and the decoding speed and queueing latency of the LLM services we use vary with service load. Elapsed time would therefore mix FlowArk's effect with fluctuations in service performance. Instead, USD cost, input tokens, and ReAct turns more directly measure paid model consumption and analysis effort.

Total E2E Cost includes the USD cost reported by OpenCode and additional history-reuse LLM costs from FlowArk's knowledge management or the Mem0 backend; local embedding computation is unpriced. Completed Tasks per \$100 expresses throughput under a fixed budget. Precision, relative recall, and relative F1 summarize the manual quality audit. In cost breakdowns, Analysis covers the task-analysis process and data-flow report generation, including their retries and repairs; knowledge management is counted separately.

\subsection{RQ1: To what extent does FlowArk reduce end-to-end token/API cost?}
\label{sec:eval-efficiency}

\noindent\textbf{Experimental design.}
To answer RQ1, we compare \emph{Standard OpenCode} and \emph{FlowArk-enabled OpenCode} on \textsc{Main49} under GLM-4.7. Standard OpenCode preserves OpenCode's native agent behavior without cross-instance knowledge reuse. FlowArk-enabled OpenCode enables all three FlowArk modules. We measure their end-to-end cost and fixed-budget throughput, then decompose FlowArk's added knowledge management cost.

\begin{figure}[!tbp]
    \centering
    \includegraphics[width=\linewidth]{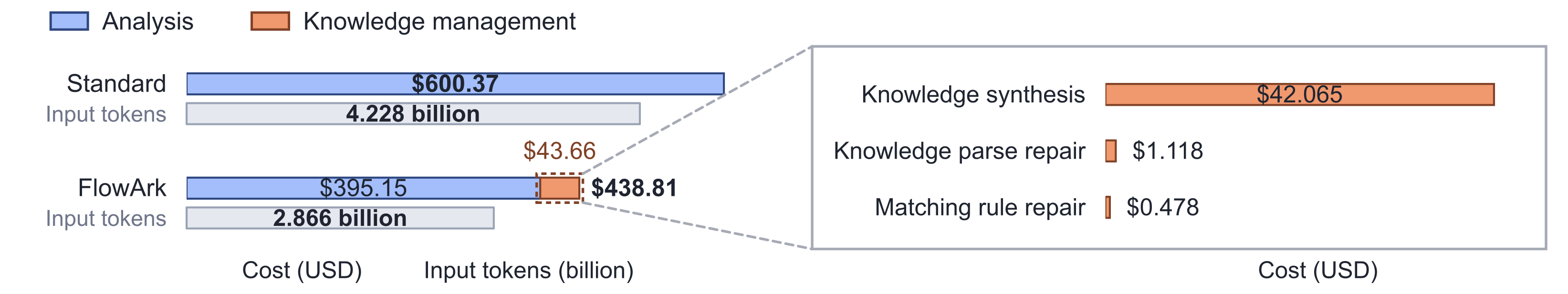}
    \vspace{-1.7em}
    \caption{Cost composition and input tokens on \textsc{Main49}. The inset details knowledge management costs.}
    \label{fig:eval-cost-composition}
    \Description{Standard costs USD 600.37. FlowArk costs USD 438.81, comprising USD 395.15 for analysis including data-flow reporting and USD 43.66 for knowledge management. The gray bars show 4.228 billion input tokens for Standard and 2.866 billion for FlowArk on a separate scale. These counts include cached inputs. Dashed lines link the management segment to an expanded inset showing USD 42.065 for knowledge synthesis, USD 1.118 for knowledge parse repair, and USD 0.478 for matching rule repair.}
\end{figure}

\noindent\textbf{End-to-end cost and throughput.}
\Cref{fig:eval-cost-composition} reports API costs and input-token counts for the 4,646 \textsc{Main49} task pairs completed in both systems. Three pairs are excluded because of execution timeouts. FlowArk-enabled OpenCode costs \$438.81, compared with \$600.37 for standard OpenCode, reducing total end-to-end cost by 26.91\%. Input-token consumption decreases from 4.228B to 2.866B, a reduction of 32.23\%. Under a USD 100 budget, the observed average cost corresponds to 1,059 completed data-flow analysis tasks instead of 774, a 36.82\% throughput increase.

\noindent\textbf{Cost composition.}
\Cref{fig:eval-cost-composition} also explains where the reduction comes from. FlowArk lowers the OpenCode analysis cost from \$600.37 to \$395.15 and adds \$43.66 for knowledge management. Because the analysis-side saving exceeds this added cost, the net end-to-end cost remains lower.

\vspace{0.5em}
\rqanswer{Answer to RQ1}{On \textsc{Main49}, FlowArk reduces input tokens by 32.23\% and end-to-end cost by 26.91\%, equivalent to 36.82\% more tasks per USD 100 at the observed average cost.}

\subsection{RQ2: How consistently does FlowArk reduce cost across repeated runs, LLM backends, and different apps?}
\label{sec:eval-stability}

\noindent\textbf{Experimental design.}
To answer RQ2, we examine cost savings across repeated runs and four LLM backends on \textsc{Sub15}, and analyze app-level variation in the \textsc{Main49} results.

We run Standard and FlowArk-enabled OpenCode three times each on the complete \textsc{Sub15} workload to assess run-to-run stability. All runs share the same task order and configuration. Each FlowArk run starts with empty knowledge bases.

Across LLM backends, we compare Standard and FlowArk-enabled OpenCode on \textsc{Sub15} under GLM-4.7, GLM-5.3~\cite{zai_pricing}, DeepSeek-V4-Flash~\cite{deepseek_pricing}, and MiniMax-M3~\cite{minimax_paygo_pricing}. For GLM-5.3, we reuse these three runs, averaging costs and paired-run cost reductions; the other backends report individual runs.

\phantomsection
\label{par:eval-repeated-stability}
\noindent\textbf{Stability across repeated runs.}
FlowArk reduces aggregate cost, input tokens, and ReAct turns in all three paired \textsc{Sub15} runs under GLM-5.3, as shown in \Cref{tab:eval-repeated-runs}. The arithmetic mean of the three per-run cost reductions is 25.73\%, with a standard deviation of 1.41 percentage points. The total-cost coefficient of variation (CV), defined as the sample standard deviation divided by the mean across runs, is 1.45\% for Standard and 1.26\% for FlowArk.

\begin{table}[t]
\centering
\caption{Cost stability across three repeated runs on \textsc{Sub15}.}
\vspace{-0.5em}
\label{tab:eval-repeated-runs}
\footnotesize
\setlength{\tabcolsep}{3pt}
\begin{tabular*}{\linewidth}{@{\extracolsep{\fill}}lrrrrrrr@{}}
\toprule
\multirow{3}{*}{\shortstack[l]{\textbf{Repeated}\\\textbf{run}}} & \multicolumn{3}{c}{\textbf{Standard}} & \multicolumn{3}{c}{\textbf{FlowArk}} & \multirow{3}{*}{\shortstack[r]{\textbf{Cost}\\\textbf{reduction (\%)}}} \\
\cmidrule(lr){2-4}\cmidrule(lr){5-7}
& \textbf{E2E cost} & \textbf{Input tokens} & \textbf{ReAct turns} & \textbf{E2E cost} & \textbf{Input tokens} & \textbf{ReAct turns} & \\
& (USD) & ($10^6$) & per task & (USD) & ($10^6$) & per task & \\
\midrule
1 & 253.32 & 535.64 & 15.05 & 186.62 & 231.21 & 8.81 & 26.33 \\
2 & 249.32 & 533.97 & 14.94 & 182.64 & 229.49 & 8.98 & 26.75 \\
3 & 246.08 & 535.09 & 15.05 & 186.72 & 233.01 & 8.81 & 24.12 \\
\midrule
\shortstack[l]{Mean\\$\pm$ SD} & \shortstack[r]{249.58\\$\pm\,3.63$} & \shortstack[r]{534.90\\$\pm\,0.85$} & \shortstack[r]{15.01\\$\pm\,0.06$} & \shortstack[r]{185.33\\$\pm\,2.33$} & \shortstack[r]{231.24\\$\pm\,1.76$} & \shortstack[r]{8.87\\$\pm\,0.10$} & \shortstack[r]{25.73\\$\pm\,1.41$} \\
\bottomrule
\end{tabular*}
\end{table}

Individual task costs vary more across runs, reflecting the variability inherent in agent execution. On the same task, an agent may choose different analysis methods or paths across runs, leading to larger task-level cost CVs. We compute each task's sample standard deviation across the three repeated runs. For Standard and FlowArk, respectively, the medians are 2.83 and 2.89 US cents, with 90th percentiles of 11.44 and 9.41 US cents. The median task-level CVs are 20.96\% and 25.13\%, with 90th percentiles of 43.58\% and 57.81\%. Their contrast with the much lower total-cost CVs suggests that fluctuations partially offset one another across the large workload. Aggregation thus substantially attenuates their relative impact on overall cost, while some variability remains in individual executions.

\noindent\textbf{Robustness across LLM backends.}
FlowArk reduces end-to-end cost across all four evaluated LLMs, spanning different generations of the GLM family and models from different providers (\Cref{tab:eval-model-comparison}). Because pricing differs across LLMs, we also report reductions in input tokens and ReAct turns as indicators of reduced analysis effort. Differences in cost savings may reflect model pricing structures, behavioral tendencies, and capabilities; isolating the contributions of these factors is beyond the scope of this evaluation.

\begin{table}[t]
\centering
\caption{Cost and analysis effort across LLM backends on \textsc{Sub15}.}
\vspace{-0.7em}
\label{tab:eval-model-comparison}
\footnotesize
\setlength{\tabcolsep}{3pt}
\begin{tabular*}{\linewidth}{@{\extracolsep{\fill}}lrrrrr@{}}
\toprule
\textbf{Model} & \shortstack{\textbf{Standard}\\\textbf{Cost (USD)}} & \shortstack{\textbf{FlowArk}\\\textbf{Cost (USD)}} & \shortstack{\textbf{Cost}\\\textbf{Reduction}} & \shortstack{\textbf{Input Token}\\\textbf{Reduction}} & \shortstack{\textbf{ReAct Turn}\\\textbf{Reduction}} \\
\midrule
GLM-4.7 & 179.72 & 130.21 & 27.55\% & 32.72\% & 21.26\% \\
GLM-5.3 & 249.58 & 185.33 & 25.73\% & 56.77\% & 40.93\% \\
DeepSeek-V4-Flash & 16.67 & 14.41 & 13.59\% & 37.85\% & 25.78\% \\
MiniMax-M3 & 62.73 & 38.60 & 38.47\% & 50.64\% & 36.08\% \\
\bottomrule
\end{tabular*}
\end{table}

\begin{figure}[!tbp]
    \centering
    \includegraphics[width=\linewidth]{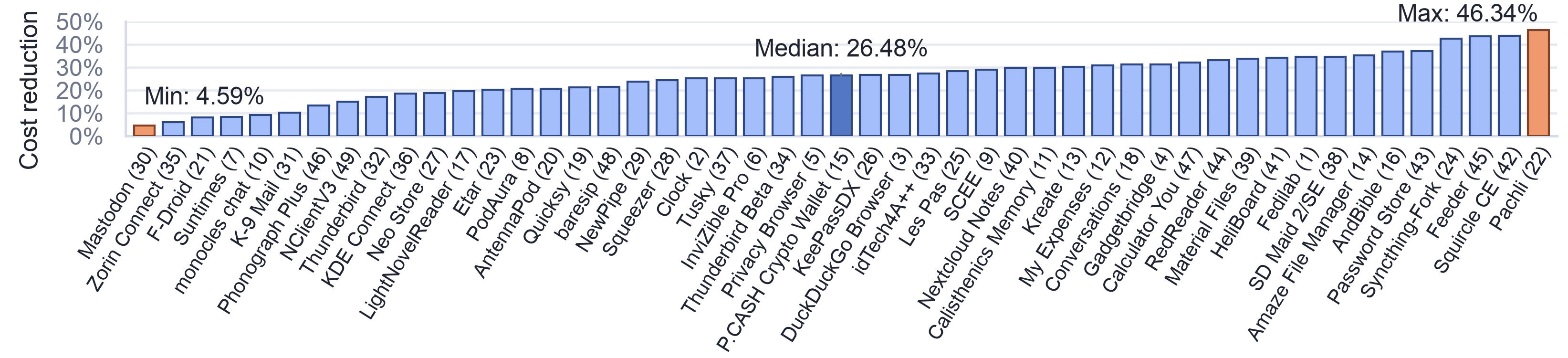}
    \vspace{-1.7em}
    \caption{Per-app E2E cost reduction of FlowArk-enabled OpenCode relative to standard OpenCode on \textsc{Main49}. Labels identify each app and its source-count rank in \textsc{Main49}.}
    \label{fig:eval-per-app}
    \Description{A bar chart compares app-level cost reductions for the 49 retained applications, with labels showing app names and source-count ranks.}
\end{figure}

\noindent\textbf{Stability across apps.}
FlowArk reduces cost across all 49 evaluated apps. \Cref{fig:eval-per-app} shows reductions from 4.59\% to 46.34\%, with a median of 26.48\%. Source-count rank has little association with app-level reduction (Pearson $0.048$, Spearman $0.043$). Source count describes analysis workload, while reuse opportunities, executed paths, and agent variation can also affect savings. Excluding any one app leaves aggregate savings between 25.93\% and 27.49\%. To check sensitivity to extreme task costs, we compute each system's 5th and 95th cost percentiles over the 4,646 completed task pairs. We retain a task pair only if both costs fall within their respective intervals. The remaining 3,789 pairs still yield 24.49\% savings. These checks show that the aggregate result persists beyond individual apps and extreme task costs.

\rqanswer{Answer to RQ2}{FlowArk reduces cost on all 49 \textsc{Main49} apps, with a median reduction of 26.48\%, and under all four LLM backends on \textsc{Sub15}. Across three repeated \textsc{Sub15} runs, the cost-reduction standard deviation is 1.41 percentage points, indicating stable workload-level savings.}

\subsection{RQ3: How does FlowArk compare with standard OpenCode and generic history-reuse mechanisms in cost and analysis quality?}
\label{sec:eval-baselines}

\noindent\textbf{Baseline selection and configuration.}
To answer RQ3, we compare FlowArk-enabled OpenCode with Standard OpenCode, Mem0-enabled OpenCode, and RAG-enabled OpenCode on \textsc{Sub15}.

Mem0-enabled OpenCode uses the official agent plugin and a self-hosted Mem0 backend built from the official Mem0 source, with default memory extraction and retrieval settings~\cite{mem0,mem0_opencode_plugin}. We self-host the Mem0 backend to track its LLM API costs and maintain a stable, controlled experimental environment. Both Mem0 and the data-flow analysis agents use GLM-4.7.

RAG-enabled OpenCode indexes earlier same-app reports and logs as text chunks~\cite{wu_repoformer_2024,wang_repomem_2026,wang_memgovern_2026}. It ranks the chunks by cosine similarity using locally deployed \texttt{Qwen3-Embedding-0.6B} embeddings~\cite{zhang_qwen3_embedding_2025,qwen3_embedding_06b}. At task start, we retrieve chunks using the task prompt, app identifier, source description, and target sink categories. We then inject the top three results.

\noindent\textbf{Manual quality audit.}
We perform one shared manual quality audit on an app-stratified 20\% sample of \textsc{Sub15} (257 tasks), covering standard OpenCode, FlowArk-enabled OpenCode, the two history-reuse baselines, and the three FlowArk ablations. Tasks are selected by sorting SHA-256 hashes of stable task keys within each app. Two master's students with Java/Kotlin Android development experience independently check findings against source code with system names hidden, resolving disagreements by discussion. Different key paths remain distinct even with the same endpoints. A shared sink category or similar API names alone are insufficient grounds for merging findings. Within each task, findings are merged only when source inspection confirms the same source occurrence, sink occurrence, and key propagation path, despite differences in wording or detail. Each merged finding is counted at most once per system.

Because exhaustive source-to-sink ground truth is unavailable for these apps, we use a \emph{reported-finding-pool audit}. For each task, the reference pool contains the distinct findings reported by at least one system and manually validated against source code. Findings that cannot be reliably judged are marked \emph{uncertain} (15 of 2,861 audited findings) and excluded from the counts. For each system, $\mathrm{TP}$ and $\mathrm{FP}$ count supported and unsupported reported findings; relative false negatives ($\mathrm{FN}_{\mathrm{rel}}$) count supported pool findings it misses. We sum these counts over all audited tasks and compute micro-averaged precision $P=\mathrm{TP}/(\mathrm{TP}+\mathrm{FP})$, relative recall $R_{\mathrm{rel}}=\mathrm{TP}/(\mathrm{TP}+\mathrm{FN}_{\mathrm{rel}})$, and relative F1 $F_{1,\mathrm{rel}}=2PR_{\mathrm{rel}}/(P+R_{\mathrm{rel}})$, reported in \Cref{tab:eval-baseline-comparison,tab:eval-ablation-comparison}. The subscript $\mathrm{rel}$ indicates that the quantity is defined relative to the shared validated finding pool.

Relative F1 combines checked precision with coverage of the validated pool to support comparison among systems. Because systems may report different valid subsets, a moderate score alone does not establish poor overall analysis quality.

\noindent\textbf{Comparison results.}
We run the systems and conduct the manual quality audit as described above, yielding the results in \Cref{tab:eval-baseline-comparison}. These results show that FlowArk achieves lower cost than Standard OpenCode and both history-reuse baselines, with similar analysis quality in the manual audit. Relative to Standard, FlowArk reduces total cost by 27.55\%. On the audited tasks, it also yields 18.82 validated findings per dollar, compared with 14.30 for Standard, a 31.64\% increase.

\begin{table}[t]
\centering
\caption{Baseline comparison on \textsc{Sub15}.}
\vspace{-0.5em}
\label{tab:eval-baseline-comparison}
\small
\setlength{\tabcolsep}{1pt}
\renewcommand{\arraystretch}{1.05}
\newcommand{\auditdelta}[1]{\ensuremath{(#1\%)}}
\newcommand{\bestauditdelta}[1]{{\boldmath\ensuremath{(#1\%)}}}
\begin{tabular*}{\linewidth}{@{\extracolsep{\fill}}l*{2}{r@{\extracolsep{0pt}\hspace{2pt}}r@{\extracolsep{\fill}\hspace{2\tabcolsep}}}rrrrrr@{}}
\toprule
\textbf{System} & \multicolumn{2}{c}{\textbf{E2E Cost (USD)} $\downarrow$} & \multicolumn{2}{c}{\textbf{Tasks/\$100} $\uparrow$} & {\boldmath$\mathrm{TP}$} $\uparrow$ & {\boldmath$\mathrm{FP}$} $\downarrow$ & {\boldmath$\mathrm{FN}_{\mathrm{rel}}$} $\downarrow$ & {\boldmath$P$} $\uparrow$ & {\boldmath$R_{\mathrm{rel}}$} $\uparrow$ & {\boldmath$F_{1,\mathrm{rel}}$} $\uparrow$ \\
\midrule
Standard & 179.72 &  & 714 &  & 505 & 394 & 608 & 56.17\% & 45.37\% & 50.20\% \\
Mem0 & 175.85 & \auditdelta{-2.16} & 730 & \auditdelta{+2.20} & \textbf{559} & 529 & \textbf{554} & 51.38\% & \textbf{50.22\%} & 50.80\% \\
RAG & 181.53 & \auditdelta{+1.01} & 707 & \auditdelta{-1.00} & 515 & \textbf{366} & 598 & \textbf{58.46\%} & 46.27\% & \textbf{51.65\%} \\
\midrule
FlowArk & \textbf{130.21} & \bestauditdelta{-27.55} & \textbf{986} & \bestauditdelta{+38.03} & 507 & 413 & 606 & 55.11\% & 45.55\% & 49.88\% \\
\bottomrule
\end{tabular*}
\par\smallskip
\raggedright\footnotesize
Cost and throughput use all 1,284 \textsc{Sub15} tasks; quality metrics use the 257 audited tasks.
\end{table}

Mem0-enabled OpenCode provides little cost benefit, reducing total cost by only 2.16\% over Standard OpenCode. Its relative F1 remains close to FlowArk's result (50.8\% vs. 49.9\%), suggesting a limited quality tradeoff alongside the larger efficiency gap. We interpret this cost gap as a result of its generic, agent-initiated memory use. Without data-flow-oriented distillation, code-anchor matching, or context-aware runtime injection, the agent may incur extra memory overhead yet fail to retrieve useful knowledge at the moment repeated analysis can be skipped.

RAG-enabled OpenCode achieves slightly higher relative F1 than FlowArk (51.7\% vs. 49.9\%), while its total cost is 1.01\% higher than Standard OpenCode. It struggles to deliver reusable evidence when the agent reaches the relevant code context. Its task-start retrieval lacks runtime matching, and reusable evidence is sparse in verbose data-flow logs. We use task-start retrieval because injecting log chunks after every tool call could incur input-token costs that outweigh the savings from reuse.

\vspace{0.5em}

\rqanswer{Answer to RQ3}{On \textsc{Sub15} under GLM-4.7, FlowArk achieves lower cost than Standard OpenCode and both evaluated history-reuse baselines, with more validated findings per dollar than Standard. The manual audit indicates similar quality on sampled tasks.}

\subsection{RQ4: What contributes to FlowArk's observed cost savings?}
\label{sec:eval-ablation}

\noindent\textbf{Experimental design.}
To answer RQ4, we examine knowledge delivery and repeated code inspection, compare injections of relevant and irrelevant knowledge to test whether cost savings stem from knowledge reuse, and ablate individual FlowArk modules.

\emph{(1) Source-line access reconstruction.}
We compare repeated source-line access counts reconstructed from execution logs to assess whether FlowArk reduces code rereading across tasks within the same app. For the 4,646 completed task pairs in \textsc{Main49}, we reconstruct source-line accesses from recorded \texttt{read} and native \texttt{grep} outputs, together with outputs of common shell commands such as \texttt{grep}, \texttt{rg}, \texttt{cat}, and \texttt{sed}. After excluding ambiguous matches and duplicates within each task, we count repeated appearances of the recovered file--line pairs across tasks in the same app. We then compare these counts between Standard OpenCode and FlowArk-enabled OpenCode.

\emph{(2) Knowledge replacement experiment.}
To test whether FlowArk's cost savings are attributable to knowledge reuse rather than to prompts or other factors, we compare Standard OpenCode and FlowArk-enabled OpenCode with \emph{FlowArk w/ irrelevant knowledge}, a variant that only replaces matched knowledge with unrelated knowledge from other apps. Replacement covers all injections: each matched block is replaced before delivery with an unrelated knowledge block of similar character length (0.8--1.25 times the original) from another app. The replacement limits useful knowledge reuse while keeping knowledge structure, prompts, and injection behavior comparable.

We repeat each condition three times for more reliable conclusions. To limit experimental cost, we use a subset of recorded \textsc{Sub15} executions. App-stratified systematic sampling covers the full task sequence as knowledge accumulates. Within each app's original execution order, a fixed hash determines a reproducible random starting position, and we select every fourth task thereafter. Excluding five candidates without historical knowledge or eligible replacements leaves 314 tasks across all 15 \textsc{Sub15} apps. Each replacement task starts in a fresh agent instance using the frozen knowledge state available before its paired FlowArk execution.

\emph{(3) Module ablations.}
The w/o M1--M3 ablations use simpler substitutes to keep FlowArk's coupled knowledge reuse pipeline operational. We compare these ablations with the full FlowArk system. The ablations use the same manual quality audit and metrics as RQ3 (\Cref{sec:eval-baselines}), covering 257 sampled tasks.

\emph{w/o M1} removes FlowArk's selection-rule constraints and his\-tor\-i\-cal-overlap-guided boundary selection. \emph{w/o M2} replaces rule-based matching with embedding-based recall of distilled knowledge. This differs from RAG-enabled OpenCode, which retrieves historical logs directly. \emph{w/o M3} recalls knowledge only from the initial task context, disabling runtime matching after tool calls.

\noindent\textbf{Observed reuse and repeated code inspection.}
Knowledge delivery reaches a large share of tasks and accompanies less repeated code inspection. For the 4,646 completed task pairs in \textsc{Main49}, FlowArk injects knowledge into the agent contexts of 4,345 tasks (93.52\%), delivering 27,054 knowledge injection blocks and sending 25,507 runtime messages.

Using the source-line access counts reconstructed above, we find that FlowArk reduces cross-task repeated source-line inspection by 34.36\%. This reduction is consistent with knowledge reuse helping agents skip previously analyzed code.

\noindent\textbf{Contribution of knowledge reuse.}
Replacing matched knowledge with irrelevant knowledge brings FlowArk's analysis effort closer to that of Standard OpenCode (\Cref{tab:eval-knowledge-reuse-contribution}). We replace matched blocks in 314 sampled tasks with similarly sized knowledge from other apps, retaining prompts and injection behavior. Analysis cost and input-token reductions relative to Standard fall from 48.25\% and 64.45\% to 7.04\% and 7.78\%, respectively. ReAct turns rise from 7.86 to 13.80 per task, approaching Standard's 15.16. Of the resulting cost increase over FlowArk, 93.28\% comes from the analysis phase. This supports effective knowledge reuse as a major source of FlowArk's analysis cost savings.

\begin{table}[htbp]
\centering
\caption{Contribution of knowledge reuse to cost savings (mean $\pm$ SD over three runs).}
\vspace{-0.5em}
\label{tab:eval-knowledge-reuse-contribution}
\small
\setlength{\tabcolsep}{4pt}
\renewcommand{\arraystretch}{1.05}
\begin{tabular*}{\linewidth}{@{\extracolsep{\fill}}lrrrrr@{}}
\toprule
\textbf{System} & \shortstack{\textbf{Input tokens}\\($10^6$)} & \shortstack{\textbf{ReAct turns}\\\textbf{per task}} & \shortstack{\textbf{Analysis}\\(USD)} & \shortstack{\textbf{Knowledge}\\\textbf{mgmt.}~(USD)} & \textbf{E2E}~(USD) \\
\midrule
Standard & $136.34 \pm 7.41$ & $15.16 \pm 0.29$ & $62.87 \pm 3.08$ & $0.00 \pm 0.00$ & $62.87 \pm 3.08$ \\
FlowArk & {\boldmath$48.47 \pm 5.88$} & {\boldmath$7.86 \pm 0.37$} & {\boldmath$32.53 \pm 2.62$} & $13.45 \pm 0.36$ & {\boldmath$45.99 \pm 2.87$} \\
\midrule
{\renewcommand{\arraystretch}{1}\begin{tabular}[c]{@{}l@{}}FlowArk w/\\\quad irrelevant knowledge\end{tabular}} & $125.74 \pm 4.09$ & $13.80 \pm 0.27$ & $58.45 \pm 1.56$ & $15.32 \pm 0.21$ & $73.77 \pm 1.72$ \\
\bottomrule
\end{tabular*}
\par\smallskip
\raggedright\footnotesize
Input tokens and ReAct turns cover only the analysis phase, excluding knowledge management.
\end{table}

\noindent\textbf{Module contributions.}
\Cref{tab:eval-ablation-comparison} compares FlowArk with its ablations on \textsc{Sub15}.

\begin{table}[tbp]
\centering
\caption{Module ablation comparison on \textsc{Sub15}.}
\vspace{-0.5em}
\label{tab:eval-ablation-comparison}
\small
\setlength{\tabcolsep}{1pt}
\renewcommand{\arraystretch}{1.05}
\newcommand{\auditdelta}[1]{\ensuremath{(#1\%)}}
\newcommand{\bestauditdelta}[1]{{\boldmath\ensuremath{(#1\%)}}}
\begin{tabular*}{\linewidth}{@{\extracolsep{\fill}}l*{2}{r@{\extracolsep{0pt}\hspace{2pt}}r@{\extracolsep{\fill}\hspace{2\tabcolsep}}}rrrrrr@{}}
\toprule
\textbf{Variant} & \multicolumn{2}{c}{\textbf{E2E Cost (USD)} $\downarrow$} & \multicolumn{2}{c}{\textbf{Tasks/\$100} $\uparrow$} & {\boldmath$\mathrm{TP}$} $\uparrow$ & {\boldmath$\mathrm{FP}$} $\downarrow$ & {\boldmath$\mathrm{FN}_{\mathrm{rel}}$} $\downarrow$ & {\boldmath$P$} $\uparrow$ & {\boldmath$R_{\mathrm{rel}}$} $\uparrow$ & {\boldmath$F_{1,\mathrm{rel}}$} $\uparrow$ \\
\midrule
FlowArk & \textbf{130.21} &  & \textbf{986} &  & \textbf{507} & 413 & \textbf{606} & 55.11\% & \textbf{45.55\%} & 49.88\% \\
\midrule
w/o M1 & 138.09 & \auditdelta{+6.05} & 930 & \auditdelta{-5.71} & 451 & 386 & 662 & 53.88\% & 40.52\% & 46.26\% \\
w/o M2 & 139.48 & \auditdelta{+7.13} & 921 & \auditdelta{-6.65} & 463 & 422 & 650 & 52.32\% & 41.60\% & 46.35\% \\
w/o M3 & 166.63 & \auditdelta{+27.97} & 771 & \auditdelta{-21.86} & 485 & \textbf{307} & 628 & \textbf{61.24\%} & 43.58\% & \textbf{50.92\%} \\
\bottomrule
\end{tabular*}
\par\smallskip
\raggedright\footnotesize
Cost and throughput use all 1,284 \textsc{Sub15} tasks; quality metrics use the 257 audited tasks.
\end{table}

\textbf{w/o M1: generic distillation.}
This variant has lower relative F1 (46.3\% vs. 49.9\%) and a 6.05\% cost increase relative to FlowArk. We attribute these differences to vague knowledge from generic distillation, which can distract the agent and prompt repeated inspection of the underlying code.

\textbf{w/o M2: embedding recall.}
This variant has lower relative F1 (46.3\% vs. 49.9\%) and a 7.13\% cost increase relative to FlowArk. Mismatches between retrieved knowledge and the active code context may contribute to these differences by distracting the agent and prompting repeated code inspection.

\textbf{w/o M3: initial-only injection.}
This variant costs 27.97\% more than FlowArk, the largest cost increase among the three variants. Initial-only injection can miss reuse opportunities revealed by later tool outputs, leaving the agent to re-analyze shared code. This highlights the importance of timely injection: matched knowledge must reach the agent when reuse opportunities emerge, before re-analysis.

\rqanswer{Answer to RQ4}{Reduced repeated code inspection and the irrelevant-knowledge experiment support knowledge reuse as the main source of FlowArk's cost savings. Context-aware runtime knowledge injection is the most influential module in the ablations, enabling timely reuse; selective knowledge distillation and knowledge packaging provide reliable knowledge for this process.}

\section{Discussion}
\label{sec:discussion}

\noindent\textbf{Practical implications.}
As coding agents move toward enterprise-scale deployment, their cost becomes increasingly difficult to ignore. FlowArk targets this cost at the workload level: instead of reducing the price of a single model call, it reduces repeated reading and reasoning across context-isolated agent instances. This allows a fixed budget to complete more data-flow analysis tasks, making agentic data-flow analysis more practical under cost-bounded deployment settings.

\noindent\textbf{Threats to internal validity.}
FlowArk relies on knowledge distillation, so an incomplete knowledge summary may omit information from the original analysis history and cause a later agent to miss part of a data-flow path. The agent may also produce an overly broad matching rule or an incorrect knowledge entry that pollutes the knowledge base. FlowArk already includes safeguards such as admission checks, matching-rule filtering, and later overwrite or repair to reduce these risks, but they cannot guarantee that every later agent instance will discount problematic knowledge that accidentally slips through these safeguards, which may still lead to unnecessary analysis branches or occasional incorrect conclusions.

\noindent\textbf{Threats to external validity.}
FlowArk is most useful when a batch workload contains recurring shared code fragments. Apps with little shared logic provide fewer reuse opportunities and may see smaller savings, as reflected by the per-app cost-reduction variance in our evaluation. Our evaluation focuses on Android data-flow analysis, so applying FlowArk to other programming languages requires further validation. Still, the same reuse principle may apply across languages: common software engineering practices, such as encapsulating frequently used logic, create shared code fragments whose analysis knowledge can be distilled and reused. We study reuse under a fixed codebase state; knowledge invalidation across revisions is outside our scope, though practical deployments could link entries to version metadata or commit hashes to reduce stale reuse.

\section{Related Work}
\label{sec:related-work}

\subsection{LLM-Assisted Program Analysis}

Recent LLM-assisted systems reason about data-flow paths, infer sources, sinks, or taint specifications, and audit repositories with tool support \cite{wang_llmdfa_2024,li_iris_2025,liu_llm-powered_2025,ghebremichael_multi-agent_2026,guo_repoaudit_2025}. Context-rich and project-scale studies, together with code-property-graph systems, highlight the need for cross-function and repository context \cite{lekssays_llmxcpg_2025,li_llm-based_2026,li_everything_2025,yildiz_benchmarking_2025}. Hybrid pipelines embed LLMs in static-analysis workflows through slices, static facts, sound analysis shells, warning refinement, query synthesis, or rule-guided review \cite{chapman_interleaving_2024,cheng_llm-enhanced_2024,wei_hallucination-resilient_2025,li_hitchhiker_2025,jaoua_combining_2025,wang_qlcoder_2025,yang_knighter_2025}. These works improve the precision, coverage, or evaluation of analysis pipelines. FlowArk complements them by addressing workload-level re-analysis across many context-isolated agent instances analyzing the same Android app.

\subsection{Knowledge Reuse in Program Analysis}

Traditional program analysis has long reused analysis artifacts to avoid repeated computation. Incremental and persistent analyses reuse data-flow results, caches, invariants, and demanded summaries across versions or sessions \cite{arzt_reviser_2014,cai_leveraging_2018,dusing_persisting_2023,razafintsialonina_reusing_2025,stein_interactive_2024}, while taint-summary and specification-inference systems make framework, library, or policy behavior reusable \cite{arzt_stubdroid_2016,staicu_extracting_2020,chibotaru_scalable_2019,chiang_inference_2024}. These artifacts are typically centered on solvers, frameworks, or libraries; however, in Android analysis, callbacks, ICC links, and framework models can still force later agent instances to repeat code-level confirmation \cite{cao_edgeminer_2015,li_iccta_2015,yan_comprehensive_2023}. FlowArk applies this principle to batch agentic data-flow analysis, reusing knowledge from completed agent instances when later instances reach a matching runtime analysis context.

\subsection{Agent Memory, Retrieval, and Runtime Reuse}

Agent memory systems preserve prior interactions or task experience for later coding assistance \cite{mem0,wang_repomem_2026,wang_memgovern_2026,deng_memcoder_2026}. Security-oriented agents also retain audit context, verified vulnerabilities, or related-flow evidence for later reasoning \cite{guo_repoaudit_2025,jiang_vecho_2026}. Studies on action granularity, code representation, and selective retrieval show that extra context affects behavior, cost, and latency \cite{yu_recode_2025,pan_hidden_2025,wu_repoformer_2024}. FlowArk specializes memory for batch agentic data-flow analysis by storing matchable knowledge entries and injecting them when the active agent reaches the corresponding shared code fragment.

\section{Conclusion}

When source-to-sink data-flow analysis tasks are executed as a batch workload, context-isolated agent instances may repeatedly analyze the same shared code fragments, causing re-analysis cost to accumulate across the workload. FlowArk reduces this cost by distilling reusable knowledge entries and injecting those that match the agent's runtime analysis context to skip re-analysis. Our evaluation shows that FlowArk reduces end-to-end cost by 26.91\%, and achieves a stronger cost-quality tradeoff than Mem0-enabled OpenCode and RAG-enabled OpenCode baselines. The cost savings persist across repeated runs, and trace analysis shows 34.36\% less cross-task repeated source-line inspection. These results show that timely, context-triggered knowledge reuse can reduce re-analysis cost across the workload.

\bibliographystyle{ACM-Reference-Format}
\bibliography{references}

\end{document}